\documentstyle[12pt]{article}
\begin{document}
\title{Radioactive Decay Seen as Temporal Canonical Ensemble}
\author{Slobodan Prvanovi\'c \\
{\it Scientific Computing Laboratory,}\\
{\it Center for the Study of Complex Systems,}\\
{\it Institute of Physics Belgrade, } \\
{\it University of Belgrade, Pregrevica 118, 11080 Belgrade,}\\
{\it Serbia}}
\date{}
\maketitle

\begin{abstract}
The operator of time formalism is applied to radioactive decay. It appears that the proposed approach offers better insight and understanding of the 
phenomena in a way that the decay exponential-law becomes the Boltzmann distribution in Gibbs treatment of canonical ensemble. The radioactive decay 
is seen as temporal canonical ensemble where the radioactive constant appears as the analog of the absolute temperature multiplied by Boltzmann 
constant. The stochastic character of decay process becomes plausible in the proposed approach and the explanation why decay is characterized by 
constant, and not by some parameter, is offered.


\end{abstract}

Perhaps the most intriguing fact about radioactive decay is that such process is stochastic in its nature. This means that we are unable to predict 
when some radioactive system (particle) will decay. We can only say what is the probability that it will decay at the moment $t$ or later, and this 
is given by the well known exponential law $e ^{-\lambda t}$, where $\lambda$ is radioactive constant characteristic for the decaying system under 
consideration [1]. The unpredictability when decay will occur is not the only thing that puzzles us regarding decay. There are questions what influences 
decay in general, how, if possible, can we alter $\lambda$ and can we prepare anyhow the systems to decay faster/slower. These topics are going to be 
addressed in the present article, and this will be done in a manner that is not, up to my knowledge, previously considered.

Let us jump to conclusion and say that it is not the Hamiltonian that governs decay, but some other observable. More concretely, the observable that is, 
so to say, dynamical analog of the time, here denoted by $\hat G = G (\hat q , \hat p )$, dictates, by its spectrum, at what moments particle can decay. 
This observable appears in equation that is the time analog of the Schr\"odinger equation [2], so it appears to be for decay what the Hamiltonian is for 
canonical ensemble.

In order to show how decay process can be treated as temporal canonical ensemble the formalism of operator of time, that we have proposed in [2-6], will 
be used. There is a whole variety of topics and approaches related to the operator of time, {\it e.g.}, [7-9] and references therein. Our approach 
is similar to [10], and references therein, and [11], and its crucial point is to treat time and energy as the coordinate and momentum are usually treated. 
This means that separate Hilbert space where operators of time and energy act is introduced, just as it is done for each degree of freedom in standard 
formulation of quantum mechanics. In this way the Pauli's objection is avoided. Then, the same commutation relation that holds for coordinate and 
momentum is imposed for energy and time, which leads to unbounded spectrum of these operators. Finally, the original and second Schr\"odinger equation, 
that we have introduced in [2], as constraints in the overall Hilbert space select physically meaningful states that have non-negative energy and time. 

As it is done for every spatial degree of freedom, a separate Hilbert space ${\cal H}_t$, where operators of time $\hat t$ and energy $\hat s$ act
non-trivially, can be introduced. So, for the case of one degree of freedom, there are $\hat q \otimes \hat I$, $\hat p \otimes \hat I$, $\hat I
\otimes \hat t$ and $\hat I \otimes \hat s$, acting in ${\cal H}_q \otimes {\cal H}_t$, and for these self-adjoint operators the following
commutation relations hold:
$$
 {1\over i\hbar} [\hat q \otimes \hat I, \hat p \otimes \hat I ] = \hat I \otimes \hat I ,
$$
$$
 {1\over i\hbar} [\hat I \otimes \hat t , \hat I \otimes \hat s ] = - \hat I \otimes \hat I .
$$
The other commutators vanish. The operators of time $\hat t$ and energy $\hat s$ have continuous spectrum $\{ -\infty , +\infty \}$, just like the 
operators of coordinate and momentum $\hat q$ and $\hat p$. The eigenvectors of $\hat t$ are $\vert t \rangle $ for every $t\in {\bf R}$. (The 
question related to the norm and measurement of $\vert t \rangle$ is analyzed in [10].) In $\vert t \rangle$ representation, operator of energy is 
given by $i \hbar {\partial \over \partial t}$ and its eigenvectors $\vert E \rangle$ in the same representation are $e^{{1\over i\hbar} Et}$ for 
every $E\in \bf R$. In [3] we have shown how the unbounded spectrum of the operator of energy is regulated by the Schr\"odinger equation. Let us here 
just stress that the Schr\"odinger equation, that appears as constraint in ${\cal H}_q \otimes {\cal H}_t$, selects physically meaningful states with 
non-negative energy, due to the bounded from below spectrum of Hamiltonian. In [2], we have introduced, {\it s. c.}, second Schr\"odinger equation:
\begin{equation}
\hat t \vert \psi \rangle = G (\hat q, \hat p ) \vert \psi \rangle  .
\end{equation}
In (1) one demands that the operator of time and its dynamical counterpart $G (\hat q, \hat p )$ equally act on states of quantum mechanical system,
just like in the original Schr\"odinger equation $\hat s \vert \psi \rangle = H (\hat q, \hat p ) \vert \psi \rangle$ one demands that the operator
of energy and the Hamiltonian, as its dynamical counterpart, equally act on states of quantum mechanical system. (That this is Schr\"odinger
equation could be verified by taking its $\vert q \rangle \otimes \vert t \rangle$ representation.) As the original Schr\"odinger equation, the
equation (1) represents constraint in ${\cal H}_q \otimes {\cal H}_t$, as well. The typical solution of (1) is $\vert \psi _t \rangle \otimes \vert t
\rangle $, where $G (\hat q , \hat p ) \vert \psi _t \rangle = t \vert \psi _t \rangle$ and $\hat t \vert t \rangle = t \vert t \rangle$. It is the
time analog of $\vert \psi _E \rangle \otimes \vert E \rangle $ that solves the original Schr\"odinger equation if $H (\hat q , \hat p ) \vert
\psi _E \rangle = E \vert \psi _E \rangle$ and $\hat s \vert E \rangle = E \vert E \rangle$. (In $\vert q \rangle \otimes \vert t \rangle$ representation,
the last state becomes $\psi _E (q)e^{{-1\over i \hbar} Et}$.)

Due to the Big Bang as the beginning of time, it seems reasonable to assume that $G(\hat q , \hat p )$ has bounded from below spectrum, just like the
Hamiltonian. In what follows, for the sake of simplicity, we shall assume that $G(\hat q , \hat p )$ and $H(\hat q , \hat p )$ have discrete eigenvalues
$t_i$ and $E_i$, and the solutions of original and second Schr\"odinger equations will be denoted as $\vert E_i \rangle \otimes \vert E_i \rangle $ and
$\vert t_i \rangle \otimes \vert t_i \rangle $, respectively.

In the proposed framework of ${\cal H}_q \otimes {\cal H}_t$, the well known statements regarding canonical ensemble are as follows. Suppose the system is 
characterized by the Hamiltonian of the form $\hat H = \sum _i E_i \vert E _i \rangle \langle E_i \vert$. The canonical ensemble is the statistical 
operator:
\begin{equation}
\hat \rho _E= {1\over Z} \sum _i e^{-\beta E_i} \vert E _i \rangle \langle E_i \vert \otimes \vert E _i \rangle \langle E_i \vert .
\end{equation}
In the last expression, the first $\vert E _i \rangle \langle E_i \vert$ is the projector on the eigenstate of Hamiltonian for eigenvalue $E_i$, that is 
$\vert E_i \rangle$ which belongs to ${\cal H}_q$, while the second one is projector on the eigenvector of $\hat s$ for the same eigenvalue $E_i$, that is 
$\vert E_i \rangle$ which belongs to ${\cal H}_t$. The canonical partition function $Z_E$ is determined by normalization condition:
\begin{equation}
Z_E= {\rm Tr} \sum _i e^{-\beta E_i} \vert E_i \rangle \langle E_i \vert \otimes \vert E_i \rangle \langle E_i \vert .
\end{equation} 
Usually, $\beta$ is seen as $k_B T$, where $k_B$ is Boltzmann constant and $T$ is the absolute temperature, but $\beta$ can be taken as temperature {\it per se}. 

The Boltzmann distribution $e^{-\beta E_i}$ is probability distribution over various states $\vert E _i \rangle \otimes \vert E _i \rangle$. According 
to the second law of thermodynamics, state of equilibrium maximizes the entropy, and maximization of entropy leads to Gibbs prescription for statistical operator 
(2), {\it i. e.}, to Boltzmann distribution. Among all ensembles with the same mean value of $\hat H$ or $\hat s$, the canonical ensemble has the maximal 
entropy. The expression relating temperature, internal energy $\langle \hat H \rangle$ and entropy $ -\langle \hat \rho ln \hat \rho \rangle$ is: 
\begin{equation}
\beta =k_B T = -{\partial \langle \hat H \rangle \over \partial \langle \hat \rho ln \hat \rho \rangle }=-{\partial {\rm Tr} \hat \rho \hat H \over \partial 
{\rm Tr } \hat \rho ln \hat \rho }.
\end{equation}
For the canonical ensemble (2) it holds:
\begin{equation}
[ \hat \rho _E, \hat s ] =0,
\end{equation}
the meaning of which is that $\hat \rho$ does not depend on time:
\begin{equation}
{\partial \hat \rho _E\over \partial \hat t }=0,
\end{equation}
since $\hat s$ is generator of time translation. 

On the other side, radioactivity is described by the well known formula:
\begin{equation}
N(t_i)=N_0 e^{- \lambda t_i}.
\end{equation}
Here, $N(t_i)$ is the number of decaying systems (particles) present at the moment $t_i$ if there were $N_0$ at the $t_0$. (For the sake of simplicity we 
shall consider time as having discrete values $t_i$.) Equivalent description of decay process is to ask what is the probability that some system will last 
until $t_i$, when it will decay, and the corresponding expression for this is:
\begin{equation}
{e^{- \lambda t_i}\over \sum _i e^{- \lambda t_i} }.
\end{equation}
If $\hat G$ has descrete spectrum, then solutions of (1) are $\vert t_i \rangle \otimes \vert t_i \rangle$. By using probability distribution (8), one can 
form statistical operator:
\begin{equation}
\hat \rho _t= {1\over \sum _i e^{- \lambda t_i} } \sum _i e^{-\lambda t_i} \vert t _i \rangle \langle t_i \vert \otimes \vert t _i \rangle \langle t_i \vert .
\end{equation}
This statistical operator represents temporal canonical ensemble that describes decay process. There is complete analogy among $\hat \rho _E$ and $\hat \rho _t$, 
the meaning of which is that as probability distribution $e^{- \beta E_i}$ determines how likely it is to find the system under consideration having the energy $E_i$, 
the probabilty distribution $e^{- \lambda t_i}$ determines how likely it is to find that the system under consideration will last until the moment $t_i$. The 
analogy is complete in sense that canonical ensemble and radioactive decay appear to be essentially the same phenomena for two conjugate observables $\hat s$ and 
$\hat t$. This does not come as surprise after noticing similarity between original and second Schr\"odinger equation and the fact that both, $\hat H$ and 
$\hat G$, that determine solutions of these equations, have bounded from below spectra. Having this in mind, instead of approaching heuristically as was done here, 
one can start with $\hat \rho _E$, change $E_i$ with $t_i$ and $\beta$ with $\lambda$, and by proceeding in reverse order arive to decay formula (7).

Obviously, in analogy with the canonical partition function $Z_E$, one can introduce its temporal counterpart:
\begin{equation}
Z_t= {\rm Tr} \sum _i e^{-\lambda t_i} \vert t_i \rangle \langle t_i \vert \otimes \vert t_i \rangle \langle t_i \vert .
\end{equation}
Moreover: 
\begin{equation}
\lambda = -{\partial \langle \hat G \rangle \over \partial \langle \hat \rho ln \hat \rho \rangle }=-{\partial {\rm Tr} \hat \rho \hat G \over \partial 
{\rm Tr } \hat \rho ln \hat \rho },
\end{equation}
so one can relate radioactive constant and mean value of $\hat G$. For instance, by using the example given in [2], if $G(\hat q , \hat p )$ is:
$$
G(\hat q , \hat p )= {\hbar \over m^2 c^4} ({\hat p^2 \over 2m} + {1\over 2} m \omega \hat q^2),
$$
then solutions of the second Schr\"odinger equation (1) are $\vert \psi _{n} \rangle \otimes \vert t_n \rangle $, $n \in {\bf {\rm N}}$, where $\vert \psi _{n} 
\rangle $ are well known solutions of the eigenvalue problem for Hamiltonian of harmonic oscillator and:
$$
t_n = {\hbar ^2 \omega \over m^2 c^4} (n + {1\over 2}).
$$ 
By introducing $d={\hbar ^2 \omega \over m^2 c^4}$, the relation among $\lambda$ and $\langle \hat G \rangle$ is then:
\begin{equation}
\langle \hat G \rangle = {1\over 2} coth ( {\lambda d \over 2}).
\end{equation}
So, this example shows that by measuring half-life one can find the mean value of $\hat G$.

From the obvious fact that commutator $[\hat \rho _t , \hat t ]$ vanishes, it follows:
\begin{equation}
{\partial \hat \rho _t\over \partial \hat s }=0,
\end{equation}
since $\hat t$ is generator of energy translation. The meaning of this equation is that the radioactive decay, seen as the temporal canonical ensemble, does 
not depend on the (internal) energy, which is well known from the experience. 

As the absolute temperature appears in $\beta =k_B T$ after the Boltzmann constant is introduced, with the appropriate constant $l$, one can define the 
persistence $P$ by:
$$
\lambda =l P .
$$
However, there is a difference between absolute temperature and persistence (or $\beta$ and $\lambda$). Namely, it is possible to change temperature of canonical 
ensemble by putting it in a contact with the other one. During thermalisation, systems exchange energy. If one system 
changes its state from the one with energy $E_1$ to the state with $E_2$, then the other system can change its state, too. If the difference in energy between the 
energy levels of the second system does not match $E_2 - E_1$, then part of the energy can come from or go to the kinetic energy of systems under consideration. 
In this process, system can instantaneously change its momentum. On the other side, similar changes of states with sharp values of time are limited by impossibility 
to exceed the speed of light. If one system changes its state from the one with duration of existence $t_1$ to the state with $t_2$, then the other system has to 
change its state in a way by exactly matching $t_2 - t_1$. Otherwise, since the coordinate is for time what the momentum is for energy (appropriate components of 
quadri vector), instantaneous change in coordinate, by some finite amount, would be necessary. Since this can contradict the fact that the velocity greater than $c$ 
is impossible, the process analog to thermalisation and changes in persistence of temporal canonical ensembles are hard to manage. In this is the reason why 
radioactive elements are characterized by radioactive constant, while $\beta$, or $T$, appearing in common canonical ensemble (2), can vary. 

The radioactive decay offers concrete example where formalism related to the second Schr\"odinger equation, that has been introduced in [2], finds its 
applicability. There we have introduced 'dynamical' counterpart of time $\hat G$ and, from the above given, its importance becomes obvious for it is for decay 
what the Hamiltonian is for standard canonical ensemble. The eigenstates of $G$ determine at which instance of time decay can happen, which is the analog of 
the eigenstates of Hamiltonian, that determine energy levels of considered system.   

In proposed formalism, the exponential decay law appears to be probability distribution characteristic to canonical ensembles. In this way, the Boltzmann 
distribution, used in Gibbs treatment of canonical ensembles, gets on its universality. The normalized Boltzmann distribution gives the probability to find 
system in some state, say $\vert t_a \rangle \langle t_a \vert \otimes \vert t_a \rangle \langle t_a \vert $. As is the case for every mixed state, {\it a priori} 
it is not possible to know the state of some particular system from the ensemble described by (9). This is why radioactive decay looks like the stochastic process. 
Namely, we do not know when some system will decay because we do not know in which state $\vert t_a \rangle \langle t_a \vert \otimes \vert t_a \rangle \langle 
t_a \vert $ the system is. We only know the probability to find the system in $\vert t_a \rangle \langle t_a \vert \otimes \vert t_a \rangle \langle t_a 
\vert $, {\it i.e.}, the probability that the system will not decay before $t_a$. In other words, radioactive decay is not essentialy different from any other 
mixed state regarding randomness and stochasticity. It is not possible to predict when some system will decay as it is not possible to be certain about the energy 
of some system that belongs to the canonical ensemble. 

\section{Acknowledgement}

We acknowledge support of the the Serbian Ministry of education, science and technological development, contract ON171017.

\end{document}